\documentclass{ptephy}

\usepackage{bm}
\usepackage{amssymb}
\usepackage{amsmath}

\usepackage{amsmath,amsfonts}
\usepackage{amssymb}
\usepackage{mathrsfs}

\def\defscript{\mathscr}

\def\M{{\defscript M}}
\def\N{{\defscript N}}

\def\P{{\defscript P}}
\def\Q{{\defscript Q}}

\def\X{{\defscript X}}

\def\ifempty#1{\def\tmpdata{#1}\ifx\tmpdata\empty }

\def\linebreak{\hfill\break}

\def\bra<#1|{\langle #1\rvert}
\def\ket|#1>{\lvert#1 \rangle}
\def\braket<#1|#2>{\langle #1|#2 \rangle}

\def\const{\text{const}}
\def\otop#1{\hbox{$#1\kern-0.1em$\llap{\hbox{\raise1.7ex\hbox{$\scriptstyle\circ$}}}} }

\def\inpare#1{\left(#1\right)}
\def\bigpare(#1){\left(#1\right)}
\def\inrbra#1{\left\{ #1 \right\}}
\def\insbra#1{\left[ #1 \right]}

\def\bigbra[#1]{\left[ #1 \right]}

\def\h{\hat }

\def\d{\dot }

\def\b{\bar }

\def\tend{\rightarrow}
\def\then{\Rightarrow\quad}
\def\equivalent{\quad\Leftrightarrow\quad}
\def\therefore{\mbox{\setbox0=\hbox{X}\hbox{$\ldotp$}\raise0.7\ht0\hbox{$\ldotp$}\hbox{$\ldotp$}} \quad }
\def\because{\mbox{\setbox0=\hbox{X}\raise0.7\ht0\hbox{$\ldotp$}\hbox{$\ldotp$}\raise0.7\ht0\hbox{$\ldotp$}}\kern0pt }

\def\upin{\hbox{\setbox0=\hbox{$\cup$} \vrule width 0.05 \wd0 height \ht0 depth 0pt \kern - 0.5\wd0 \box0 }}
\def\Frac(#1/#2){\left(\frac{#1}{#2}\right)}

\def\Mat#1{\begin{pmatrix} #1 \end{pmatrix}}

\def\Tr{{\rm Tr}}

\def\sdprod{\mathrel{{\setbox0=\hbox{$\displaystyle\times$}\lower0.3\wd0\hbox{$\stackrel{\box0}{\scriptstyle\sim}$}}}}

\def\tosigma#1,{%
    \ifx\tmpindex\relax \def\tmpindex{#1} \let\next=\tosigma
    \else \ifnum\tmpindex=0 1 \else \sigma_\tmpindex \fi
          \ifx#1\relax  \let\next=\relax
          \else \otimes \let\next=\tosigma \def\tmpindex{#1} \fi
    \fi \next}
\def\tspb(#1){\let\tmpindex=\relax\tosigma#1,\relax,}
\def\pd{\partial}

\def\HyperG(#1,#2;#3;#4){F\inpare{\textstyle #1,#2;#3;#4}}

\def\VHB{{\mathbb V}}
\def\SHB{{\mathbb S}}

\def\Eq#1{\begin{equation} #1 \end{equation}}

\def\Eqr#1{\begin{eqnarray} #1 \end{eqnarray}}

\def\Eqrsub#1{\begin{subequations}\Eqr{#1}\end{subequations}}
\def\Eqrsubl#1#2{\begin{subequations}
  \expandafter\ifx\csname Rlabel\endcsname \relax \label{#1}
  \else \Rlabel{#1} \fi \Eqr{#2}\end{subequations}}
\def\Bitm{\begin{itemize}}
\def\Eitm{\end{itemize}}
\def\Blist#1#2{\begin{list}{#1}{\parsep=0pt \itemsep=0pt%
  \listparindent=0pt #2}}
\def\Elist{\end{list}}
\long\def\ignore#1#2{\def\ignoreflag{#1}\long\def\tmptext{#2}
  \ifnum\ignoreflag>1 #2 \fi}

\renewcommand{\titlepagefootline}{\relax}
\renewcommand{\titlepageheadline}{\relax}

\begin{document}

\title{Stability of the Schwarzschild-de Sitter black hole in the dRGT massive gravity theory}

\author{ Hideo Kodama$^{1,2}$ and Ivan Arraut$^{1,3}$}

\address{\affil{1}{Theory Center, Institute of Particle and Nuclear Studies, KEK Tsukuba, Ibaraki, 305-0801, Japan}
\affil{2}{Department of Particle and Nuclear Physics,
Graduate University for Advanced Studies, Tsukuba 305-0801, Japan}
\affil{3}{Department of Physics, Osaka University, Toyonaka, Osaka 560-0043, Japan}
}

\begin{abstract}
The Schwarzschild-de Sitter solution in the Einstein theory with a positive cosmological constant $\Lambda=m^2/\alpha$ becomes an exact solution to the dRGT non-linear massive gravity theory with the mass parameter $m$ when the theory parameters $\alpha$ and $\beta$ satisfy the relation $\beta=\alpha^2$.  We study the perturbative
behaviour
of this black hole solution in the non-linear dRGT theory with $\beta=\alpha^2$. We find that the linear perturbation equations become identical to those for the vacuum Einstein theory when they are expressed in terms of the gauge-invariant variables.
This implies that this black hole is stable in the dRGT theory as far as the spacetime structure is concerned
in contrast to the case of the bi-Schwarzschild solution in the bi-metric theory.  However, we have also found a pathological feature that the general solution to the perturbation equations contain a single arbitrary function of spacetime coordinates.  This implies a degeneracy of dynamics
in the St\"uckelberg field sector
at the linear perturbation level in this background.
Physical significance of this degenercy depends on how the St\"uckelberg fields couple observable fields.
\end{abstract}

\maketitle

\section{Introduction}

One of the biggest problems in cosmology is to explain the current accelerated expansion of the universe. In the standard theory of gravity, i.e. general relativity, this reduces to the cosmological constant ($\Lambda$) problem or the dark energy problem \cite{DEproblem,RC15} if we require the spatial homogeneity(Cf. \cite{Tomita.K2001,Kodama.H&Saito&Ishibashi2010,Goto.H&Kodama2011}). Beside this standard approach, many alternative theories have been suggested in order to solve this problem. Among the most populars, we have modified gravity theories (MOG) \cite{Moffat}, non-localities \cite{S. Deser} and massive gravity theories \cite{dRGT}, which are just large scale modifications of gravity.

In order for such a theory to be a real theory of nature, it must be consistent with all the observed features. In particular, it must be consistent with the 'observed' existence of astrophysical black holes. In many case, this requirement leads to non-trivial constraints. For example, it was recently claimed that the bi-Schwarzschild solution is unstable against a spherically symmetric perturbation in the bi-metric theory of gravity\cite{Babichev.E&Fabbri2013}. Motivated by this, the stability of the Schwarzschild-de Sitter black hole was analyzed in the framework of the linear massive gravity theory by Brito, Cardoso and Pani\cite{Brito.R&Cardoso&Pani2013,Brito.R&Cardoso&Pani2013a}.  They found that the black hole is unstable generically, but becomes stable when the mass of the graviton takes the particular value $m^2=2\Lambda/3$. In this case, the theory is inside the regime of partially massless gravity, where the Vainshtein mechanism seems to be unnecessary since the DVZ discontinuity does not appear anymore\cite{PMG}. However, it has been demonstrated that the partially massless theories of gravity have several problems of consistency \cite{S. Deser2}.

In the present paper, we analyze the stability of the Schwarzschild-de Sitter solution in the framework of the non-linear dRGT massive theory of gravity. We do not introduce the cosmological constant as an extra parameter of the theory, but instead, we utilize the fact that the Schwarzschild-de Sitter black hole is an exact solution to the non-linear dRGT theory if the parameters $\alpha=1+3\alpha_3$ and $\beta=3(\alpha_3+4\alpha_4)$ of the theory satisfy the relation $\beta=\alpha^2$. For this parameter choice, the mass term of the theory behaves exactly as the cosmological constant term in the Einstein theory for a spherically symmetric geometry as pointed out by Berezhiani et al\cite{Berezhiani.L&&2012}.  We exhaust all Schwarzschild-de Sitter-type solutions to the non-linear dRGT theory in the unitary gauge for the St\"uckelberg fields assuming $\beta=\alpha^2$. We find a family of solutions that are gauge equivalent to the standard Schwarzschild-de Sitter solution if we neglect the non-trivial transformation of the St\"uckelberg fields.  In the massive gravity theory, they should be regarded as different solutions because if the metric are put into the standard Schwarzschild-de Sitter form, the St\"uckelberg fields behave differently.The solution obtained in \cite{Berezhiani.L&&2012} is one solution in this family that is regular at the future horizon. There exists no solution that is regular both at the future and the past horizons.

We consider linear perturbations of this background solution in the framework of the nonlinear dRGT theory only assuming the parameter relation $\beta=\alpha^2$. Hence, we generally expect to obtain perturbation equations that are different from those in the Einstein theory with the cosmological constant. In fact, we do if we do not impose the constraint coming from the Bianchi identity on the mass term. However, when we impose that constraint, the extra terms are required to vanish. Hence, we obtain the perturbation equations that are identical to those in the Einstein theory with a cosmological constant and some additional constraints on the metric perturbation variables that correspond to the gauge-dependent parts in the Einstein theory.  From this result and the Birkhoff theorem for the Einstein theory, we can easily find the general solution to the perturbation equations and deduce the stability of the black hole against linear perturbations
concerning the spacetime structure.  However, we
also find that this general solution contains an arbitrary function of the spacetime coordinates
that reduces to a part of the gauge transformation freedom in the absense of the St\"uckelberg fiels.
In the gauge in which the background metric takes the standard Schwarzschild-de Sitter form, this freedom goes to the St\"uckelberg fields. Hence, we cannot determine the behavior of the fields only by initial data. Along with this general argument, we point out that the general solution to the vector-type  perturbation equations contains a family of stationary modes that correspond the rotation of a black hole in the Einstein theory.

The paper is organized as follows. In Section \ref{sec:dRGT}, we summarize the basic part of the dRGT non-linear massive gravity formalism that is relevant to the present paper. In Section \ref{sec:BHsol}, we show that the mass term in the field equation of the dRGT theory becomes identical to the cosmological constant term for an arbitrary spherically symmetric metric in the unitary gauge for the St\"uckelberg fields when the theory parameters satisfy the relation $\beta=\alpha^2$, and that as a consequence the Schwarzschild-de Sitter spacetime becomes an exact solution in the dRGT theory for this parameter relation. We also develop details of the Schwarzschild-de Sitter solution in the dRGT theory.  In Section \ref{sec:GIFBHreview}, we make a brief review of the gauge-invariant formulation for perturbations of a black hole. In Section \ref{sec:PA}, we derive  perturbation equations for the Schwarzschild-de Sitter type background in the dRGT theory, and then in Section \ref{sec:GIFdRGT}, we introduce gauge-invariant variables for the present system by treating the St$\ddot{u}$ckelberg fields to be dynamical and express the perturbation equations in terms of them. In Section \ref{sec:Conc}, we summarize and conclude. In Appendix \ref{App:TheOtherCases}, we show that there exist other parameter choices for which the dRGT theory admits a Schwarzschild-de Sitter type solution and exhaust all possibilities.

\section{The dRGT theory}   \label{sec:dRGT}

In the standard formalism of the dRGT theory, the action is given by \cite{dRGT}
\Eq{   \label{MG:action}
S=\frac{1}{2\kappa^2}\int d^4x\sqrt{-g}(R+m^2U(g,\phi))
}
with the effective potential depending on two free parameters as
\Eq{   \label{U:def}
U(g,\phi)=U_2+\alpha_3 U_3+\alpha_4U_4.
}
The dependence of each term $U_n$ on the metric $g$ and the St\"uckelberg field $\phi^a$ is determined in terms of the matrix $\Q=(Q^\mu{}_\nu)$ defined by
\Eqrsub{
&& \Q=1 -\M,\quad (\M^2)^\mu{}_\nu=g^{\mu\lambda} f_{\lambda\nu},\label{Q-Matrix:def}\\
&& f_{\mu\nu}=\eta_{ab} \pd_\mu\phi^a \pd_\nu \phi^b, \label{f-metric:def}
}
as
\Eqrsub{
&& \label{b3}
U_2=Q_1^2-Q_2,\\
&&\label{b4}
U_3=Q_1^3-3Q_1Q_2+2Q_3,\\
&& \label{b5}
U_4=Q_1^4-6Q_1^2Q_2+8Q_1Q_3+3Q_2^2-6Q_4,\\
}
where
\Eq{
 \label{b6}
Q_n=\Tr(\Q^n).
}
The potential $U$ is unique. It is impossible to add polynomial terms without introducing a ghost \cite{dRGT,Berezhiani.L&&2012}.

By taking a variation of the action with respect to the metric, we obtain the field equation
\Eq{
\label{MG:FieldEq}
G_{\mu \nu}=-m^2X_{\mu\nu},
}
where
\Eq{
\label{X-Matrix:def}
X_{\mu \nu}=\frac{\delta U}{\delta g^{\mu \nu}}-\frac{1}{2}Ug_{\mu \nu}.
}
Its mixed components $\X=(X^\mu{}_\nu)=g^{\mu\lambda}X_{\lambda\nu}$ can be explicitly expressed in the matrix form in terms of the matrix $\Q$ as
\Eq{
\X = \chi_0 + \chi_1 \Q + \chi_2  \Q^2 + \chi_3 \Q^3,
\label{XbyQ}
}
where
\Eqrsub{
\chi_0 &=&-\frac{\beta}{3}Q_3+\frac{\alpha+\beta Q_1}{2}Q_2 -Q_1 -\frac{\alpha}{2}Q_1^2-\frac{\beta}{6}Q_1^3,\\
\chi_1 &=& 1 + \alpha Q_1 + \frac{\beta}{2}(Q_1^2-Q_2),\\
\chi_2 &=& -\alpha-\beta Q_1,\\
\chi_3 &=& \beta,
}
with
\Eq{
\alpha=1+3\alpha_3,\quad
\beta=3(\alpha_3+4\alpha_4)
}
Throughout the present paper, we use $\alpha$ and $\beta$ instead of $\alpha_3$ and $\alpha_4$.

In this generally covariant formulation, we can regard the St\"uckelberg fields either to be dynamical or to be non-dynamical.  This is because the dynamical equation for $\phi^a$ obtained from the action by a variation with respect to $\phi^a$ is practically equivalent to the consistency equation obtained from \eqref{MG:FieldEq} by the Bianchi identity.

To see this, we use the diffeomorphism invariance of the mass term of the action,
\Eq{
\int d^4x'\sqrt{-g'}U(g',\phi')=\int d^4x\sqrt{-g}U(g,\phi).
}
For an infinitesimal coordinate transformation
\Eq{
\delta x=\zeta^\mu,\quad \delta g_{\mu \nu}=-2\nabla_{(\mu}\zeta_{\nu)},\quad\delta\phi=-\zeta^\mu\partial_\mu\phi,
}
this equation leads to
\Eq{
0=\int d^4x\sqrt{-g}\left(-m^2\nabla_\nu X^{\mu\nu}\zeta_\mu-\frac{\delta U}{\delta\phi}\nabla_\mu\phi\zeta^\mu\right).
}
Because $\zeta^\mu$ is an arbitrary vector field, we obtain
\Eq{
m^2\nabla_\nu X^\nu{}_{\mu}=-\partial_\mu\phi^a\frac{\delta U}{\delta\phi^a}=\partial_\mu\phi^a\nabla_\nu\left(\frac{\delta U}{\partial(\partial_\nu\phi^a)}\right).
}
Therefore, if the field equation \eqref{MG:FieldEq} holds, the left-hand side of this equation should vanish due to the Bianchi identity $\nabla_\nu G^\nu_\mu\equiv0$. Because $\pd_\mu \phi^a$ is a regular matrix, this constraint is equivalent to the Euler equation for the St\"uckelberg field,
\Eq{
\nabla_\mu\left(\frac{\partial U}{\partial(\partial_\mu\phi^a)}\right)=0.
}

\section{The Schwarzschild-de Sitter solution}   \label{sec:BHsol}

If the Schwarzschild-de Sitter solution satisfies the field equations in massive gravity, the tensor $X_{\mu \nu}$ becomes a constant multiple of $g_{\mu \nu}$ for that metric\cite{Berezhiani.L&&2012}:
\Eq{   \label{EinsteinCondition}
m^2X_{\mu\nu}=\Lambda g_{\mu \nu}.
}
Conversely, if a solution to the field equations \eqref{MG:FieldEq} satisfies this relation, it must be a solution to the vacuum Einstein equations with $\Lambda$. Hence, if it is spherically symmetric, the solution must be diffeomorphic to the Schwarzschild-de Sitter solution. Note that this does not implies the uniqueness of the solution because although the matrices of two solutions are related by a coordinate transformation, the St\"uckelberg fields may not be related by the same transformation.

In this section, we examine under what conditions \eqref{EinsteinCondition} holds for spherically symmetric spacetimes. In particular, we show that if the parameters $\alpha_3$ and $\alpha_4$ satisfy the relation
\Eq{   \label{alpha-beta-relation}
\beta=\alpha^2,
}
any spherically symmetric metric of the form
\Eq{   \label{SSmetric}
ds^2=g_{tt}(t,r)dt^2+2g_{tr}(t,r)dtdr+g_{rr}(t,r)dr^2+r^2S(t,r)^2d\Omega_2^2
}
satisfies the condition (\ref{EinsteinCondition}) with
\Eq{   \label{LambdaBym}
\Lambda=m^2\frac{1-S_0}{S_0}=\frac{m^2}{\alpha},
}
if $S(t,r)$ is a constant given by
\Eq{   \label{Sbyalpha}
S= S_0:=\frac{\alpha}{\alpha+1}.
}

Note that the cosmological constant $\Lambda$ is different from zero for any finite value of $\alpha$ if $m^2\neq0$.

To prove this, we work in the unitary gauge in which the St\"uckelberg fields $\phi^a$ are given by
\Eq{
\phi^0=t,\quad \phi^1=x=r\cos\theta, \quad \phi^2=y=r\sin\theta\cos\phi,\quad \phi^3=z=r\sin\theta\sin\phi
\label{UnitaryGauge}
}
in the Cartesian Minkowski coordinates. In this gauge, the reference metric $f_{\mu\nu}$ in the spherical coordinates is given by
\Eq{
f_{\mu\nu}dx^\mu dx^\nu = -dt^2 + dr^2 + r^2(d\theta^2+\sin^2\theta\,d\phi^2).
}
Hence, for the metric \eqref{SSmetric}, the matrix $\M^2$ defined by \eqref{Q-Matrix:def} is given by
\Eq{   \label{M2byg}
\M^2 = \begin{pmatrix}
-g^{tt}&g^{tr}&0&0\\
-g^{tr}&g^{rr}&0&0\\
0&0&\frac{1}{S^2}&0\\
0&0&0&\frac{1}{S^2}
\end{pmatrix}.
}
From this, we find that the matrix $\Q$ can be expressed in the form
\Eq{   \label{Qbyabc}
\Q =\begin{pmatrix}
a&c&0&0\\
-c&b&0&0\\
0&0&1-\frac{1}{S}&0\\
0&0&0&1-\frac{1}{S}
\end{pmatrix},
}
where $a$, $b$ and $c$ are expressed in terms of the metric coefficients as
\Eqrsubl{abcbyg}{
&&
   1-a=\frac{1}{M_1}(-g^{tt}+(-g_{(2)})^{-1/2}), \\
&&
   c=-\frac{g^{tr}}{M_1},\\
&&
    1-b=\frac{1}{M_1}(g^{rr}+(-g_{(2)})^{-1/2}),
}
with
\Eqr{
&&
M_1=(-g_{(2)})^{-1/2}\left(-g_{tt}+g_{rr}+2(-g_{(2)})^{1/2}\right)^{1/2},\\
&&
g_{(2)}=g_{tt}g_{rr}-g_{tr}^2,
}
%

%
%

We can also express $g_{\mu \nu}$ in terms of the components of $\Q$ as
\Eqrsubl{gbyabc}{
&& g_{tt}=-\frac{(1-b)^2-c^2}{[(1-a)(1-b)+c^2]^2},\\
&& g_{rr}=\frac{(1-a)^2-c^2}{[(1-a)(1-b)+c^2]^2},\\
&& g_{tr}=-\frac{c(2-a-b)}{[(1-a)(1-b)+c^2]^2},\\
&& g_{\theta\theta}=r^2S^2,\quad
   g_{\phi\phi}=r^2S^2\sin^2\theta.
}
In particular,
\Eq{
\label{detgbyabc}
(-g_{(2)})^{-1/2}=c^2+(1-a)(1-b).
}

If we substitute the expression for  $\Q$ in terms of $a$, $b$, $c$ and $S$ into \eqref{XbyQ}, we get
\Eqrsubl{Xbyabc}{
&&
X^t{}_{t}=-bF_3-(F_1+1)\frac{(S-1)}{S},\\
&&
X^t{}_{r}=cF_3,\\
&&
X^t{}_{t}-X^r{}_{r}=(a-b)F_3,\\
&&
X^t{}_{t}-X^\theta{}_{\theta}=F_1\left(a-1+\frac{1}{S}\right)+F_2\left(ab+c^2-b\frac{(S-1)}{S}\right),
}
where $F_1$, $F_2$ and $F_3$ are functions of $S$ defined by
\Eqrsubl{Fn:def}{
&&
F_1=\alpha+1-\frac{\alpha}{S},\\
&&
F_2=\alpha+\beta-\frac{\beta}{S},\\
&&
F_3=F_1+\frac{(S-1)}{S}F_2.
}

Now, it is easy to see that all of $F_1$, $F_2$ and $F_3$ vanish if the relations \eqref{alpha-beta-relation} and \eqref{Sbyalpha} hold. This means that $\X=(X^\mu{}_{\nu})$ becomes a multiple of the unit matrix:
\Eq{
X^\mu{}_{\nu}=\frac{1-S}{S}\delta^\mu {}_{\nu}.
}
Note that this holds independent of the functional dependences of $a(t,r)$, $b(t,r)$ and $c(t,r)$.

If we require that the metric \eqref{SSmetric} be a solution of the field equations \eqref{MG:FieldEq} with \eqref{alpha-beta-relation}, owing to the Birkhoff theorem for the Einstein vacuum system, it must be isomorphic to the Schwarzschild-de Sitter solution in the standard form for which $g_{tt}=-f(r)$, $g_{tr}=0$ and $g_{rr}=1/f(r)$ with $f(r)=1-2M/r-\Lambda r^2/3$. The above result means that
$g_{tt}$, $g_{tr}$ and $g_{rr}$ obtained from this standard form by arbitrary change of time coordinate $t \tend T(t,r)$ also satisfies the field equations \eqref{MG:FieldEq}. Because we have already fixed the spacetime coordinates by the unitary gauge condition \eqref{UnitaryGauge}, these solutions obtained from the standard form by fixing the St\"uckelberg fields and applying the coordinate transformation only to the metric should be regarded to be inequivalent mutually.

Finally, we notice that the above parameter relation is not the only case in which a metric isomorphic to the Schwarzschild-de Sitter solution satisfies the field equation \eqref{MG:FieldEq}. In the Appendix \ref{App:TheOtherCases}, we exhaust all such possibilities.

\section{Gauge invariant formulation for Black Hole perturbations}   \label{sec:GIFBHreview}

In this section, we introduce some notions to describe perturbations of a black hole spacetime and its gauge-invariant treatment formulated previously in \cite{3,4}. We start from a general spherically symmetric background metric given by
\Eq{
ds^2=g_{\mu\nu}dx^\mu dx^\nu =g_{ab}(y)dy^ady^b+r^2(y)d\Omega^2,
\label{BG:metric:general}
}
where $g_{ab}$ is the metric of a two-dimensional spacetime $\N^2$ and
\Eq{
d\Omega^2=\gamma_{ij}dz^idz^j =d\theta^2+ \sin^2\theta d\phi^2
}
is the metric of a unit two-sphere $S^2$, whose  Ricci tensor is given by
$
\hat{R}_{ij}= \gamma_{ij}.
$

We denote the covariant derivative, connection coefficients and curvature tensors as
\Eq{
\nabla_\mu; \quad \Gamma^\mu_{\nu\lambda},\quad R_{\mu\nu\lambda\sigma}
}
for the four-dimensional whole spacetime,
\Eq{
D_a; \quad \Gamma^a_{bc},\quad R_{abcd}
}
for the two-dimensional spacetime $\N^2$, and
\Eq{
\hat{D}_i;\quad \hat{\Gamma}^i_{jk},\quad \hat{R}_{ijkl}=\gamma_{ik}\gamma_{jl}-\gamma_{il}\gamma_{jk}
}
for the 2-sphere $S^2$.

The spherical symmetry of the background requires the  background energy-momentum tensor to be given by
\Eq{
T_{ab}=T_{ab}(y),\quad
T_{ai}=0,\quad
T^i{}_{j}=P(y) \delta^i{}_{j}.
}

\subsection{Tensorial decomposition of perturbations}

We classify perturbation variables into two different types according to their tensorial behavior on $S^2$ so that we get a decoupled closed set of differential equations for each type of perturbations. For that purpose, we decompose the tensors $h_{ab}(y)$, $h_{ai}(y)$ and $h_{ij}(y)$ on $S^2$ defined by the metric perturbation $h_{\mu\nu}=\delta g_{\mu\nu}$ as
\Eq{
 h_{\mu\nu}dx^\mu dx^\nu= h_{ab}dy^ady^b+2h_{ai}dy^adz^i+h_{ij}dz^idz^j
}
into these irreducible tensorial components as follows.

First,  $h_{ab}$ are scalar with respect to transformations over $S^2$. Next, the vector $h_{ai}$ on $S^2$ can be uniquely decomposed into the scalar $h_a$ and the divergence-free vector $h_{ai}^{(1)}$ as
\Eq{
h_{ai}=\hat{D}_ih_a+h_{ai}^{(1)};\quad \h D^i h_{ai}^{(1)}=0,
}
up to the addition of arbitrary functions only of $y$ to $h_a$, which correspond to the exceptional $l=0$ mode (S-mode) in the harmonic expansion explained later. This implies that this exceptional mode for $h_a$ is a spurious mode and should be discarded.

Finally, the 2-tensor $h_{ij}$ on $S^2$ can be decomposed into three parts as
\Eq{
h_{ij}=2\hat{D}_{(i}h^{(1)}_{T\,j)}+h_L\gamma_{ij}+\hat{L}_{ij}h_T^{(0)};\quad
\hat{D}^ih_{T\,i}^{(1)}=0,
}
where
\Eq{
\h L _{ij} = \h D_i \h D_j - \frac12 \gamma_{ij} \h \triangle.
}
For this decomposition, $h_T^{(0)}$ is uniquely determined up to functions belonging to the kernel of the operator $\h L_{ij}$, which consists of the S-mode ($l=0$) and the $l=1$ modes in the harmonic expansion. Similarly, $h^{(1)}_{T\,i}$ is unique up to a combination of the Killing vector of $S^2$ with arbitrary functions of $y$ as coefficients. This corresponds to the exceptional mode with $l=1$ in the harmonic expansion. These exceptional modes are spurious as the S-mode for $h_a$ and should be discarded in physical arguments.  With this understanding, the scalar components $(h_{ab}, h_a ,h_L, h_T)$ of the metric perturbation $h_{\mu\nu}$ describe {\em the scalar perturbation}, and the vector components  $(h_{ai}^{(1)}, h_{T\, i}^{(1)})$ describe {\em the vector perturbation}.

In a similar way, we can decompose the energy-momentum perturbations as
\Eqrsub{
&& \delta T^a_i =\hat{D}_i\delta T^a+\delta T^{(1)a}_i;\quad \hat{D}^i\delta T^{(1)a}_i=0,\\
&& \delta T^i_j=\delta T^{(1)i}_j +\delta P \delta^i_j +\hat{L}^i_j\delta T_T^{(0)},
}
where
\Eq{
\delta T^{(1)j}_j=0,\quad
\hat{D}^j T^{(1)i}_j=0.
}
Hence, the scalar and vector components of the perturbation of the energy-momentum tensor consist of $(\delta T_{ab},\delta T^a, \delta P, \delta T^{(0)}_T)$ and $(\delta T^{(1)a}_i, \delta T^{(1)i}_j)$, respectively. There exist spurious exceptional modes in $\delta T^a$ and $\delta T_T^{(0)}$ as in the metric perturbation decomposition.

\subsection{Gauge invariant variables}

The Einstein equations are invariant under the diffeomorphism generated by any vector field $\zeta^M$. The perturbation variable $h_{\mu\nu}$ and its image $h_{\mu\nu}-\pounds_\zeta g_{\mu\nu}$ obtained by an infinitesimal diffeomorphism should represent the same physical situation. Then, we have an ambiguity since there are infinite varieties of values for the perturbation variables representing the same physical situation. One way to remove this redundancy is to construct gauge-invariant variables  and express the perturbation equations in terms of them. This automatically extracts the physical degrees of freedom related to the perturbations.

We start from the gauge transformation laws for perturbation variables.  First, for the infinitesimal gauge transformation $\delta x^\mu=\zeta^\mu$,  the metric perturbation $h_{\mu\nu}$ transforms as
\Eqrsub{
&& h_{ab}\to h_{ab}-D_a\zeta_b-D_b\zeta_a,\\
&& h_{ai}\to h_{ai}-r^2D_a\left(\frac{\zeta_i}{r^2}\right)-\hat{D}_i\zeta_a,\\
&& h_{ij}\to h_{ij}-2\hat{D}_{(i}\zeta_{j)}-2r(D^ar)\zeta_a\gamma_{ij}.
}
Next, the perturbation of the energy-momentum tensor, $\delta T_{\mu\nu}$, transforms as
\Eqrsub{
&& \delta T_{ab}\to\delta T_{ab}-\zeta^cD_cT_{ab}-T_{ac}D_b\zeta^c-T_{bc}D_a\zeta^c,\\
&& \delta T^a_i\to\delta T^a_i-T^a_b \hat{D}_i\zeta^b + P \h D_i \zeta^a,\\
&& \delta T^i_j\to\delta T^i_j-\zeta^d D_a P \delta^i_j.
}

These transformation laws can be translated to those for the perturbation variables describing each type of perturbations by decomposing the vector field $\zeta^\mu$  into vector and scalar components as
\Eq{
\zeta_a=T_a,\quad
\zeta_i=V_i+\hat{D}_iS; \quad\gamma^{ij}\hat{D}_iV_j=0.
}
Now, we execute this translation and construct gauge-invariant variables for each type of perturbations.

\subsubsection{Vector perturbations}

For vector perturbations, the above gauge transformation law for the metric perturbation can be translated into the irreducible vector components as
\Eqrsub{
&& h_{ai}^{(1)}\to h_{ai}^{(1)}-r^2D_a\left(\frac{V_i}{r^2}\right),\\
&& h_{Ti}^{(1)}\to h_{Ti}^{(1)}-V_i.
}
From this, it follows that the combination

\Eq{
F_{ai}^{(1)}=h_{ai}^{(1)}-r^2D_a\left(\frac{h_{T\,i}^{(1)}}{r^2}\right)
\label{GI:V:Fai:def}
}
is gauge invariant for generic modes. On the other hand, for the exceptional mode, $h_{T\,i}^{(1)}$ does not exist, and only the combination
\Eq{
F^{(1)}_{abi}:= 2r^2 D_{[a}\inpare{r^{-2}F_{b]i}^{(1)}}
}
is gauge invariant.

In contrast to the metric perturbation, $\delta T^a_i$ and $\delta T^i_j$ for a vector perturbation of the energy-momentum tensor become gauge invariant by themselves :
\Eqrsub{
&& \tau^{(1)a}_i:=\delta T^{(1)a}_{i}, \label{GI:V:tauai:def}\\
&& \tau^{(1)i}_j:=\delta T^{(1)i}_j. \label{GI:V:tauTi:def}
}
For the exceptional perturbations, $\tau^{(1)i}_j$ does not exist.

Note that any gauge-invariant variable for a generic vector perturbation can be expressed as a linear combination of $(F_{ai}^{(1)}, \tau^{(1)a}_i, \tau^{(1)i}_j)$ and their derivatives. Further,  we can express the perturbation variables $(h_{ai}^{(1)}, \delta T^{(1)a}_i, \delta T^{(1)i}_j)$ in terms of these three gauge-invariant variables and $h_{Ti}^{(1)}$. Under gauge transformations, $h_{Ti}^{(1)}$ just transforms like $\zeta_i$. Hence, if we express this variable in terms of the gauge-invariant variables, gauge is automatically specified.  The exceptional perturbations should be treated with more care.

\subsubsection{Scalar perturbations}

For scalar perturbations, the scalar components of the metric perturbation transform as
\Eqrsub{
&& h_{ab}\to h_{ab}-2D_{(a}T_{b)},\\
&& h_a\to h_a-T_a-r^2D_a\left(\frac{S}{r^2}\right),\\
&& h_L\to h_L-2r(D^ar)T_a-\hat{\Delta}S,\\
&& h_T\to h_T-2S.
}
If we define $X_\mu=(X_a, X_i=\hat{D}_iX_L)$ as

\Eq{
X_a:=-h_a+\frac{r^2}{2}D_a\left(\frac{h_T}{r^2}\right),\quad
X_L:=-\frac{h_T}{2},
\label{GI:S:X:def}
}
$X_\mu$ just transforms like $X_\mu \to X_\mu+\zeta_\mu$:
\Eq{
(X_a, X_L)\to(X_a+T_a,X_L+S).
}
Hence, we can define the following set of gauge-invariant variables for a generic metric perturbation:
\Eqrsub{
&& F^{(0)}_{ab}=h_{ab}+2D_{(a}X_{b)},
\label{GI:S:F0ab:def}\\
&& F^{(0)}=h_L+2r(D^ar)X_a+\hat{\Delta}X_L.
\label{GI:S:F0:def}
}
For the exceptional modes, these are not gauge invariant.

Similarly, for generic matter perturbations, we can construct the following basic gauge-invariants:
\Eqrsub{
&& \Sigma^{(0)}_{ab} =\delta T_{ab}+X^cD_cT_{ab}+T_{ac}D_bX^c+T_{bc}D_aX^c,
\label{GI:S:Sigma0ab}\\
&& \Sigma^{(0)a}_{i}=\hat{D}_i\delta T_a+T^a_b \hat{D}_iX^b -P \h D_i X^a,
\label{GI:S:Sigma0ai}\\
&& \Sigma_L^{(0)}=\delta P + X^aD_aP,
\label{GI:S:Sigma0L}\\
&& \Pi^{(0)}=\delta T^{(0)}_T.
\label{GI:S:Pi0ij}
}
For the exceptional modes, all or some of these are not gauge invariant. Further, for the S-modes, $\Sigma^{(0)}_{ai}$ and $\Pi^{(0)}_{ij}$ do not exist, and for the exceptional modes with $l=1$, $\Pi^{(0)}_{ij}$ does not exist.

As in the vector case, any gauge invariant for generic scalar perturbations can be expressed as a combination of the variables $(F^{(0)}_{ab},F^{(0)},\Sigma^{(0)}_{ab},\Sigma^{(0)a}_{i}, \Sigma_L^{(0)},\Pi^{(0)})$ and their derivatives. Further, when we express the metric and matter perturbation variables in terms of these gauge invariants and $X_\mu$,  we can fix gauge by specifying the $X_\mu$ as a linear function of the gauge-invariant variables.  In the next section, we work in the unitary gauge to derive perturbation equations for the Schwarzschild-de Sitter black hole in the dRGT theory, and then in Section \ref{sec:GIFdRGT}, we will express the perturbation equations obtained  there in the gauge-invariant form using the  formulation explained here.

\subsection{Harmonic expansions}

In practical arguments, it is often more convenient to use the harmonic expansions for perturbation variables and their gauge-invariant combinations. We also use it in the subsequent sections. So, we here give some expressions for scalar and vector  harmonic expansions relevant to the analysis in our paper, but more details can be found in \cite{3,4}.

First, in order to expand vector perturbations, we use the irreducible harmonic vectors defined by the eigenvalue problem
\Eq{
\h\triangle \VHB_i=-k_v^2 \VHB_i,\quad \h D_i \VHB^i=0.
}
For $S^2$, the eigenvalue $k_v^2$ is given by
\Eq{
k_v^2=l(l+1)-1,\quad l=1,2,\cdots.
}
Note that $\VHB_i$ is proportional to $\epsilon_{ij}\h D^j \SHB$ where $\SHB$ is some scalar harmonics with the same $l$.
The lowest mode with $l=1$ is exceptional because it can be shown to be a Killing vector field on $S^2$ and satisfies
\Eq{
\VHB_{ij} := -\frac{1}{k_v}\h D_{(i}\VHB_{j)}=0.
}
The basic variables for vector perturbations can be expanded in terms of the vector-type harmonic basis as
\Eq{
h^{(1)}_{ai}=rf_a \VHB_i,\quad
h^{(1)}_{T\,i}=-\frac{r^2}{k_v}H_T \VHB_i,
}
and correspondingly, the gauge-invariant variables are expanded as
\Eq{
F_{ai}^{(1)}=rF_a \VHB_i,\quad
\tau^{(1)a}_i=r\tau^a \VHB_i,\quad
\tau^{(1)i}_j=\tau_T \VHB^i_j,
}
for the case of generic modes satisfying $m_V:=k_v^2-1=(l+2)(l-1)>0$, where the indices of the harmonic tensors are lowered and raised by $\gamma_{ij}$. Here and in the following, we omit the index for the harmonic basis and the corresponding summation symbols for simplicity.

For the exceptional modes with $m_V=0$, i.e. $l=1$, there is only one gauge-invariant:
\Eq{
F_{ab\,i}^{(1)}=r F_{ab}^{(1)}\VHB_i;\quad
F_{ab}^{(1)}=rD_a\left(\frac{F_b}{r}\right)-rD_b\left(\frac{F_a}{r}\right).
}

For scalar perturbations, we use a basis for the scalar harmonic functions satisfying the eigenvalue problem
\Eq{
\h\triangle \SHB=-k_s^2\SHB;\quad k_s^2=l(l+1),\ l=0,1,2,\cdots,
}
and the associated vector and tensors defined by
\Eq{
\SHB_i =-\frac{1}{k_s} \h D_i \SHB,\quad
\SHB_{ij}=\frac1{k_s^2} \h L_{ij} \SHB.
}
In terms of these harmonic tensors, the perturbation variables for scalar perturbations can be expanded as
\Eqrsub{
&& h_{ab}=f_{ab}\SHB,\quad
   h_a=-\frac{r}{k_s} f_a \SHB,\\
&& h_L= 2r^2 H_L \SHB,\quad
   h_T = 2\frac{r^2}{k_s^2} H_T \SHB,\\
&& \delta T_{ab}=\tau_{ab}\SHB,\quad
   \delta T^a = -\frac{r}{k_s} \tau^a \SHB,\\
&& \delta P=  \tau_L\SHB,\quad
   \delta T^{(0)}_T=\frac{r^2}{k_s^2} \tau_T \SHB,
}
and the corresponding gauge-invariant variables are
\Eqrsub{
&& F^{(0)}_{ab}=F_{ab}\SHB, \quad
   F^{(0)}=2r^2F\SHB, \\
&& \Sigma_{ab}^{(0)}=\Sigma_{ab} \SHB,\quad
   \Sigma^{(0)a}_i=r\Sigma^a \SHB_i\\
&& \Sigma_L^{(0)}=\Sigma_L \SHB,\quad
   \Pi^{(0)} = \frac{r^2}{k_s^2} \tau_T \SHB.
}
For exceptional modes, $\tau_T$ does not exist for the $l=0$ and $l=1$ modes, and $\Sigma_a$ does not exist for the $l=0$ modes.

\section{Perturbation analysis in the dRGT formalism}   \label{sec:PA}

In this section, we derive perturbation equations for the Schwarzschild-de Sitter solution in the dRGT theory with non-linear mass terms.

\subsection{Background solution}

As we have shown in Section \ref{sec:BHsol}, when the theory parameters $\alpha$ and $\beta$ satisfy the relation \eqref{alpha-beta-relation}, the Schwarzschild-de Sitter solution in the form \eqref{SSmetric} with $S=S_0$ becomes an exact solution to the field equations of the dRGT theory in the unitary gauge \eqref{UnitaryGauge} for the St\"uckelberg fields $\phi^a$. In this form of the solution, the extra constant factor $S_0$ appears in front of the angular part of the metric. In studying perturbations of this background, we remove this constant factor by the coordinate transformation $S_0 r \tend r$ so that we can use various formula for perturbations in the literature:
\Eq{
ds^2= g_{ab}(y)dy^ady^b + r^2 (d\theta^2+\sin^2\theta d\phi^2),
\label{SSmetricNew}
}
where the index $a$ and $b$ run over $0$ and $1$ with $y^0=t$ and $y^1=r$.  This coordinate transformation transforms the unitary gauge condition \eqref{UnitaryGauge} on the St\"uckelberg field to
\Eq{
\phi^0=t,\quad \phi^1=x=\frac{r}{S_0}\cos\theta, \quad \phi^2=y=\frac{r}{S_0}\sin\theta\cos\phi,\quad \phi^3=z=\frac{r}{S_0}\sin\theta\sin\phi
\label{UnitaryGaugeNew}
}
in the Cartesian Minkowski coordinates, and the reference metric $f_{\mu\nu}$ to
\Eq{
f_{\mu\nu}dx^\mu dx^\nu = -dt^2 + \frac{dr^2}{S_0^2} + \frac{r^2}{S_0^2}(d\theta^2+\sin^2\theta\,d\phi^2).
}

The metric \eqref{SSmetricNew} should be obtained from the standard form for the Schwarzschild-de Sitter solution
\Eq{
ds^2= -f(r)dt^2+ \frac{dr^2}{f(r)} + r^2 d\Omega_2^2;\quad
f(r)=1-\frac{2M}{r} - \frac{\Lambda}{3}r^2
}
by a coordinate transformation $t \tend T_0(t,r)$ where $T_0(t,r)$ is an arbitrary function of $t$ and $r$ with $\pd_t T_0\neq0$. Hence,
\Eq{
g_{tt}=-f(r)(\pd_t T_0)^2,\quad
g_{tr}=-f(r) \pd_t T_0 \pd_r T_0,\quad
g_{rr}=-f(r)(\pd_r T_0)^2 + 1/f(r).
\label{SdSsol:general}
}
Thus, the background solution has a degeneracy represented by an arbitrary function of $t$ and $r$ even under the spherical symmetry requirement. This degeneracy cannot be gauged away because of the existence of the St\"uckelberg fields. This implies that the dRGT theory is dynamically pathological at this background. We will see that this degeneracy extends to freedom represented by an arbitrary function of full coordinates in the linear perturbation level.

The above $r$-coordinate rescaling also affects the $\Q$ matrix. Because the dRGT theory has general covariance, $(g^*f_*)=(g^{\mu\alpha}f_{\alpha\nu})$ transforms as
\Eq{
g^*f_* \tend T^{-1}{g^*}{f_*}T;\quad
T=\Mat{ 1 & 0 & 0 & 0\\ 0 & 1/S_0 & 0 & 0 \\ 0 & 0 & 1 & 0\\ 0 & 0 & 0 & 1}
}
Because the mixed tensor $\Q$ should behave exactly as $g^*f_*$ under a coordinate transformation, the $r$-rescaling transforms $\Q$ from the old value ${\Q}'$ to
\Eq{   \label{Q-matrix:new}
\Q= T^{-1}{\Q}' T=\begin{pmatrix}
a&\frac{c}{S_0}&0&0\\
-S_0 c&b&0&0\\
0&0&1-\frac{1}{S_0}&0\\
0&0&0&1-\frac{1}{S_0}
\end{pmatrix}
}

Note that due to the $r$-rescaling, the expression for  $g_{ab}$  in terms of $a$, $b$, and $c$ is modified as follows:
\Eqrsubl{gbyabc:new}{
&& g_{tt}=-\frac{(1-b)^2-c^2}{[(1-a)(1-b)+c^2]^2},\\
&& S_0^2g_{rr}=\frac{(1-a)^2-c^2}{[(1-a)(1-b)+c^2]^2},\\
&& S_0 g_{tr}=-\frac{c(2-a-b)}{[(1-a)(1-b)+c^2]^2},\\
&& S_0^{-1}(-g_{(2)})^{-1/2}=c^2+(1-a)(1-b).
}
Similarly, $a$, $b$ and $c$ are expressed in terms of the new metric $g_{ab}$ as
\Eqrsubl{abcbyg:new}{
&&
   1-a=\frac{1}{\b M_1}(-S_0 g^{tt}+(-g_{(2)})^{-1/2}), \\
&&
   c=-\frac{g^{tr}}{\b M_1},\\
&&
    1-b=\frac{1}{\b M_1}(S_0^{-1}g^{rr}+(-g_{(2)})^{-1/2}),
}
with
\Eqr{
&&
\b M_1=(-g_{(2)})^{-1/2}\left(-g_{tt}+S_0^2 g_{rr}+2S_0(-g_{(2)})^{1/2}\right)^{1/2},\\
&&
g_{(2)}=g_{tt}g_{rr}-g_{tr}^2.
}

\subsection{Perturbation of $\X$}

Now, we calculate the perturbation of the tensor $\X=(X^\mu_\nu)$ corresponding to the metric perturbation
\Eq{
h_{ab}=f_{ab}(t,r) Y,\quad
h_{ai}=r f_a (t,r) Y_i,\quad
h_{ij}=2r^2 \insbra{ H_L Y \gamma_{ij} + H_T Y_{ij}},
}
where $Y$, $Y_i$ and $Y_{ij}$ represents the corresponding tensors for either the scalar or vector harmonics. For vector perturbations, the terms in proportion to $Y$ do not exist.

First, from \eqref{XbyQ}, a perturbation of the matrix $\X$ is determined by $\delta \Q$ as
\Eqr{
\delta \X &=& \delta\chi_0+\delta \chi_1 \Q + \delta\chi_2 \Q^2  \notag\\
           && + \chi_1 \delta \Q + \chi_2 \delta \Q^2 + \chi_3 \delta \Q^3.
}
Here, $\delta \chi_n$ is a linear combination of $\delta Q_n$, which is given by
\Eq{
\delta Q_n = \frac {n}{2} \Tr\insbra { h^*_* \Q^{n-1}(1-\Q)},
}
where $h^*_*$ is the matrix notation for the mixed tensor $h^\mu_\nu$.

In general, $\delta \Q$ is determined as the solution to
\Eq{
(1-\Q)\delta \Q + \delta \Q (1-\Q)= -\delta (\M^2)= h^*_* \M^2.
\label{eq:deltaQ:general}
}
In solving this, it is important that the background metric $g$ and the matrix $\M=g^* f_*$ are the direct sum of two-dimensional submatrices,
\Eqrsub{
&& g=g_{(1)}(t,r) \oplus g_{(2)}(\theta,\phi),\\
&& \M=\M_{(1)} \oplus \M_{(2)},
}
because the calculations of $\delta Q^a_b$, $\delta Q^a_i$ and $\delta Q^i_j$ decouple from each other except for the calculation of $\delta Q_n$, which can be directly calculated by the above formula. The results for $\delta Q_n$ are given in the Appendix \ref{App:delQn}.

First, the angular part $\delta Q^i_j$ can be easily calculated because $1-\Q_{(2)}=(1/S_0) I_2$:
\Eq{
\delta Q^i_j= \frac12 S_0 h^i_k (\M^2)^k_j = \frac1{S_0} (H_L Y \delta^i_j+ H_TY^i_j).
}
The corresponding components of $\delta \X$ are expressed in terms of this as
\Eqr{
\delta X^i_j &=& \inrbra{\delta \chi_0 + \delta\chi_1\inpare{1-\frac1{S_0}}+\delta\chi_2\inpare{1-\frac1{S_0}}^2}\delta^i_j
\notag\\
&& + \inrbra{\chi_1+2\chi_2\inpare{1-\frac1{S_0}}+3\chi_3\inpare{1-\frac1{S_0}}^2}\delta Q^i_j.
}
The result of the calculation is
\Eq{
\delta X^i_j = w(r) (H_L\delta^i_jY - H_T Y^i_j),
}
where
\Eqr{
w(r) &=& \frac{1+\alpha}{\alpha} \inrbra{ \beta(c^2+ab)+ \alpha(a+b)+1}.
}

Next, for the $t-r$ part, solving the matrix equation
\Eq{
(\delta^a_c-Q^a_c)\delta Q^c{}_b + \delta Q^a{}_c (\delta^c_b-Q^c_b)= -\delta (\M^2)^a{}_b=f^a{}_c (\M^2)^c_b Y,
}
we obtain
\Eqr{
\delta \Q_{(1)} &=& -\frac1{2(2-\Tr\Q_{(1)})}\insbra{
  \delta (\M^2_{(1)}) + \det(1-\Q_{(1)})(1-\Q_{(1)})^{-1}\delta (\M^2)_{(1)}(1-\Q_{(1)})^{-1} }
  \notag\\
  &=& \frac{1}{2(2-a-b)}\insbra{h^*_* \M^2_{(1)}+ \inrbra{c^2+(1-a)(1-b)}
   (1-\Q_{(1)}) f^{**} h_{**} (1-\Q_{(1)}) }.
}
Inserting this into
\Eq{
\delta X^a{}_b = \delta\chi_0 \delta^a_b+ \delta\chi_1 Q^a_b + \delta\chi_2 (\Q^2)^a_b
 + \chi_1\delta Q^a{}_b + \chi_2 (\delta \Q^2)^a{}_b+ \chi_3(\delta \Q^3)^a{}_b,
}
we find
\Eq{
\delta X^a{}_b=0.
}

Finally, because $(\Q^n)^a{}_i=0$ for the background $\Q$, we have
\Eq{
\delta  X^a{}_i = \chi_1 \delta Q^a{}_i + \chi_2 \delta (\Q^2)^a{}_i + \chi_3 \delta (\Q^3)^a{}_i.
}
Here,
\Eqrsub{
\delta (\Q^2)^a{}_i &=& (1-1/S_0) \delta Q^a{}_i + Q^a{}_b \delta Q^b{}_i,\\
\delta (\Q^3)^a{}_i &=& (1-1/S_0)^2 \delta Q^a{}_i + (1-1/S_0)Q^a_b{}\delta Q^b{}_i + (\Q^2)^a{}_b \delta Q^b{}_i.
}
Hence,
\Eqr{
\delta X^a{}_i &=& \inrbra{\chi_1+(1-1/S_0)\chi_2 + (1-1/S_0)^2 \chi_3}\delta Q^a{}_i
\notag\\
&& + \inrbra{\chi_2+(1-1/S_0)\chi_3}Q^a{}_b \delta Q^b{}_i
  + \chi_3 (\Q^2)^a{}_b \delta Q^b{}_i.
}
Now, \eqref{eq:deltaQ:general} for $\delta Q^a{}_i$ reduces to
\Eq{
\insbra{(1+1/S_0)\delta^a_b- Q^a{}_b} \delta Q^b{}_i =\frac{r}{S_0^2} f^a Y_i.
}
Hence, we obtain
\Eqr{
\delta X^a{}_i &=& \frac1{S_0^2} \Big[
\inrbra{\chi_1+(1-1/S_0)\chi_2 + (1-1/S_0)^2 \chi_3}\delta^a_b \notag\\
&& + \inrbra{\chi_2+(1-1/S_0)\chi_3}Q^a{}_b
  + \chi_3 (\Q^2)^a{}_b \Big] \notag\\
&&\quad \times \inpare{[1+1/S_0-\Q_{(1)}]^{-1}}^b{}_c f^c Y_i.
}
By inserting the above background value for $\Q_{(1)}$, we find this vanishes identically!!:
\Eq{
\delta X^a{}_i=0.
}

\subsection{Vector perturbations}

For vector perturbations, the metric perturbation $h_{\mu\nu}=\delta g_{\mu\nu}$ has the harmonic expansion
\Eq{
h_{ab}=0,\quad
h_{ai}=r f_a \VHB_i,\quad
h_{ij}=2r^2 H_T \VHB_{ij}.
}
Similarly, a vector perturbation of the energy-momentum tensor
\Eq{
\kappa^2\tau^\mu_\nu:=\kappa^2\delta T^\mu_\nu = -m^2 \delta X^\mu_\nu
}
has the harmonic expansion
\Eq{
\tau^a_b=0,\quad
\tau^a_i = r \tau^a \VHB_i,\quad
\tau^i_j= \tau_T \VHB^i_j,
}
where $\tau^a$ and $\tau_T$ are gauge-invariant.

From the calculations in the previous section, we obtain
\Eqrsub{
&& \tau^a=0,\\
&& \kappa^2\tau_T=m^2w(r)H_T.
}
These source terms have to satisfy the Bianchi identities, which for a vector perturbation reduce to\cite{3,4}
\Eq{
D_a(r^3\tau^a)+\frac{(l+2)(l-1)}{2[l(l+1)-1]^{1/2}}r^2 \tau_T=0
\then (l-1)w(r) H_T=0.
}
Because $w(r)\neq0$ for $\beta=\alpha^2$, it follows that $H_T=0$ for $l\ge2$. Hence, the perturbation equations are identical to those for the vacuum Einstein system, and for $l\ge2$, we obtain the additional constraint $H_T=0$. This implies that the general solution to the perturbation equation is given by
\Eq{
f_a= F_a,\quad H_T=0
}
where $F_a$ is the gauge-invariant variable for vector perturbations satisfying the perturbed vacuum Einstein equations
\Eqrsub{
&& \frac1{r^3}D^b\inpare{r^3F^{(1)}_{ab}} -\frac{m_v}{r^2}F_a=-2\kappa^2\tau_a=0,\\
&& \frac{k_v}{r^2} D_a(rF^a) = -\kappa^2\tau_T=0.
}
In particular, we can conclude that the system is stable for vector perturbations.

For the exceptional mode with $l=1$ for which $H_T$ does not exist, $F_a$ is not gauge-invariant and transforms for $\zeta^a=0, \zeta^i= L \VHB^i$ as
\Eq{
\delta F_a=-r D_a L.
}
We know that the general solution for $l=1$ in the Einstein case is a linear combination of this gauge mode and the rotational perturbation corresponding to the angular momentum component in the Kerr metric\cite{4}.  Hence, the general solution in the present case is given by
\Eq{
f_a=-r D_a L -\frac{2aM}{r}\pd_a T_0(t,r).
}
In particular, this shows that the dRGT theory admit a rotational black hole solution in the linear perturbation level.

\subsection{Scalar perturbations}

For scalar perturbations,
\Eq{
\delta X_{ab}=\delta g_{ac} X^c_b + g_{ac}\delta X^c{}_b= \frac{\Lambda}{m^2}f_{ab} \SHB.
}
Hence, the perturbation of the effective energy-momentum tensor is given by
\Eqrsub{
&& \tau_{ab}= -\Lambda f_{ab},\\
&& \kappa^2\tau^a=0,\\
&& \kappa^2\delta P=-m^2 w(r) H_L,\\
&& \kappa^2 \tau_T=m^2w(r)H_T.
}
The corresponding standard gauge-invariant variables are
\Eqrsub{
&& \kappa^2\Sigma_{ab}=\kappa^2 \tau _{ab}-2\Lambda D_{(a}X_{b)}=-\Lambda F_{ab},\\
&& \kappa^2\Sigma_a = \kappa^2 \tau_a =0,\\
&& \kappa^2\Sigma_L =-m^2w H_L
}
and $\tau_T$. These should satisfy the conservation laws\cite{3,4}
\Eqrsub{
&& \frac1{r^3}D_a(r^3\Sigma^a) -\frac{k_s}{r} \Sigma_L + \frac{k_s^2-2}{2k_sr}\tau_T=0,\\
&& \frac{1}{r^2}D_b\insbra{r^2(\Sigma^b_a+\Lambda F^b_a)}+\frac{k_s}{r}\Sigma_a
 -2\frac{D_ar}{r}\Sigma_L=0,
}
where $k_s^2=l(l+1)$. These reduce to
\Eqrsub{
&& -2l(l+1)H_L=(l+2)(l-1)H_T\quad (l\ge1),\\
&& H_L=0.
}
Hence, for all modes including the case $l=0,1$ for which $H_T$ does not exist, we obtain the constraint $H_L=H_T=0$, and the perturbation equations are identical to those for the vacuum Einstein system with $\Lambda$, which has the structure
\Eqrsub{
&&E_{ab}=2\kappa^2 \Sigma_{ab}=0,\\
&& E^a=2\kappa^2 \Sigma^a=0,\\
&& E_L = 2\kappa^2 \Sigma_L=0,\\
&& -\frac{k_s^2}{r^2}F^a_a= 2\kappa^2\tau_T=0,
}
where $E_{ab}$, $E_a$ and $E_L$ are tensors written as differential linear combinations of the gauge-invariants $F_{ab}$ and $F$. In particular, no instability occurs. The general solution for $l\ge2$ is expressed in terms of the gauge-invariant quantities satisfying the perturbations equations for the vacuum Einstein system with $\Lambda$ as
\Eqrsub{
&& f^r=k F,\\
&& f_{ab}= F_{ab} -\frac1{k}[D_a(rf_b)+D_b(rf_a)],\quad (k>0)\\
&& H_L=H_T=0,
}
where $f^t(t,r)$ is left as an arbitrary function. This corresponds to the freedom associated with the infinitesimal coordinate transformation, $\delta t=T^t \SHB$, $\delta r=0$, $\delta z^i=0$:
\Eq{
\delta_g f_{ab}=-D_aT_b-D_b T_a,\quad
\delta_g f_a=\frac{k}{r}T_a,\quad
\delta_g H_L=\delta_g H_T=0.
}

The exceptional modes with $l=0$, $1$ should be treated with care. First, for the S-mode with $l=0$, the variables $f_a$ and $H_T$ do not exists. Hence,
\Eq{
F_{ab}=f_{ab},\quad H_L=0.
}
Now, $F_{ab}$ is not gauge invariant, and transforms as
\Eqrsub{
&& \delta_g f_{ab}=-D_a T_b-D_b T_a,\\
&& \delta_g H_L= -\frac{1}{r}T^r=0.
}
The residual gauge freedom is represented by $\delta t=T^t(t,r)$. This result is consistence with the existence of the degeneracy represented by the single function $T_0(t,r)$ in the background solution.

Because the solution satisfies the Einstein equations, from the Birkhoff theorem, we know that the general solution is a linear combination of the above gauge transformation from the background solution and the perturbation corresponding to the variation of the mass parameter in the background metric,
\Eqrsub{
&& f_{tt}=\delta M \pd_M g_{tt},\\
&& f_{rr}=\delta M \pd_M g_{rr},\\
&& f_{tr}=\delta M \pd_M g_{tr},\\
&&H_L=0.
}

Next, for the  $l=1$ mode, there exists no $H_T$ again, but now we have $f_a$. However, due to the absence of $H_T$, $F$ and $F_{ab}$ are not gauge invariant, and transforms under $\delta y^a=T^a \SHB$ and $\delta z^i= L(t.r) \SHB^i$ as
\Eqrsub{
&& \delta_g F=-\frac{k}{2}L-\frac{r}{k}g^{ra}D_a L,\\
&& \delta_g F_{ab}= -\frac1{k}[D_a(r^2 D_b L) + D_b(r^2 D_a L)].
}
$L$ is restricted by the condition $H_L=0$ as
\Eq{
\delta_g H_L=-\frac{k}{2}L - \frac1{r}T^r=0.
}
Because we know that the corresponding solutions with $l=1$ to the vacuum Einstein system is exhausted by $(F,F_{ab})$ obtained from the trivial solution $(0,0)$ by the above gauge transformation\cite{3}, the general solution to our perturbation equations with $l=1$ is given by
\Eqrsub{
&& f_{ab}= -D_a T_b - D_b T_a,\\
&& f_a = -r D_a L + \frac{k}{r}T_a,\\
&& H_L=0,
}
where
\Eq{
L= -\frac{2}{kr}T^r.
}
\section{Gauge-invariant formulation for perturbations in the dRGT theory}  \label{sec:GIFdRGT}

Because the dRGT theory is a completely general covariant theory if the St\"uckelberg field is treated as a dynamical one, the perturbation equations can be also written in the gauge-invariant form by introducing gauge-invariant variables for the perturbation of the St\"uckelberg field $\phi^\alpha$.

Let us denote a perturbation of $\phi^\alpha$ as
\Eq{
\sigma^\alpha=\delta \phi^\alpha.
}
then, from the general theory, its gauge transformation under the coordinate transformation $\delta_g x^\mu=\zeta^\mu$ is given by
\Eq{
\delta_g \sigma^\alpha = -\pounds_\zeta \phi^\alpha= -\zeta^\mu \pd_\mu \phi^\alpha.
}
In the unitary gauge, the background value of $\phi^\alpha$ is
\Eq{
\phi^t = t,\quad \phi^r=\frac{r}{S_0},\quad
\phi^\theta=\frac{\theta}{S_0}.\quad
\phi^\varphi=\frac{\varphi}{S_0}.
}
Hence, for $\delta_g y^a=T^a, \delta_g z^i=L Y^i$, $\sigma^a$ transforms as
\Eqrsub{
&& \delta_g \sigma^t=-\frac{T^t}{\mu},\quad
   \delta_g \sigma^r =-\frac{T^r}{S_0},\\
&& \delta_g \sigma_T= -\frac{L}{S_0},
}
where
\Eq{
\sigma^i = \sigma_T Y^i.
}

\subsection{Vector perturbations}

For vector perturbations, we have
\Eq{
\sigma^a=0,\quad
\sigma^i=\sigma_T\VHB^i.
}
From
\Eq{
\delta_g f_a= -rD_a L,\quad \delta_g H_T= k L,
}
we can construct a gauge-invariant variable
\Eq{
\h \sigma_T=\sigma_T + \frac{1}{kS_0} H_T
}
for generic modes with $l\ge2$, in addition to the standard gauge-invariant variable for the metric,
\Eq{
F_a= f_a+ \frac{r}{k}D_a H_T.
}
Then, the source term for the massive gravity equation can be expressed in terms of it as
\Eq{
\tau^a=0,\quad
 \kappa^2\tau_T = m^2 w(r)kS_0 \h \sigma_T.
}
Hence, in terms of the gauge-invariant $\h \sigma_T$, our result is expressed as
\Eq{
\h\sigma_T=0 \quad (l\ge2).
}
This implies that the dynamical degree of freedom of the St\"uckelberg field is completely suppressed, and the perturbation of the metric behaves exactly in the same way as for the Einstein gravity.

For the exceptional modes with $l=1$, we only have a single gauge-invariant quantity
\Eq{
\h F_a= f_a- S_0 r\pd_a \sigma_T.
}
Our analysis showed that for $l=1$, the general solution for $F_a$ is given by
\Eq{
\h F_a =-r D_a L -\frac{2\alpha M}{r}\pd_a T_0,
}
where $L(t,r)$ is an arbitrary function and $\alpha$ is an arbitrary constant corresponding to the angular momentum parameter. Thus, a functional degeneracy appears.

\subsection{Scalar perturbations}

For generic modes ($l\ge2$) of scalar perturbations, we adopt the gauge-invariant variables for $\sigma^\alpha$ defined by
\Eqrsub{
&& \h \sigma^t=\sigma^t + \frac{X^t}{\mu},\\
&& \h \sigma^r =\sigma^r + \frac{X^r}{S_0},\\
&& \h\sigma_T =\sigma_T + \frac{1}{kS_0} H_T.
}
In terms of these, the source terms corresponding to $\delta\X$ are expressed as
\Eqrsub{
&& \kappa^2\Sigma_{ab}=-\Lambda F_{ab},\\
&& \kappa^2\Sigma_a=0,\\
&& \kappa^2 \Sigma_L = m^2w(r)\inpare{\frac{kS_0}{2}\h \sigma_T + \frac{S_0}{r} D_a r \h \sigma^a -F},\\
&& \kappa^2 \tau_T = m^2w(r) kS_0\h \sigma_T.
}
We have found that all of these gauge-invariant source terms vanish, hence
\Eq{
\h \sigma^r= \frac{r}{S_0}F,\quad
\h\sigma_T=0,
}
but $\h\sigma^t(t,r)$ can be an arbitrary function. Hence, the functional degeneracy appears even for generic modes.

For the exceptional modes with $l=1$, $F$, $F_{ab}$, $\h \sigma^a$ and $\h \sigma_T$ are not gauge invariant because we have to set $H_T=0$ in their definitions and transform as
\Eqrsub{
&& \delta_g F=-\frac{k}{2}L-\frac{r}{k}D^r L,\\
&& \delta_g F_{ab}= -\frac1{k}\inrbra{ D_a(r^2D_b L)+D_b(r^2 D_aL)},\\
&& \delta_g \h \sigma^t= -\frac{r^2}{\mu k} D^t L,\\
&& \delta_g \h \sigma^r= -\frac{r^2}{S_0 k} D^t L,\\
&& \delta_g \h \sigma_T=-\frac{L}{S_0}.
}
However, we can construct the following basic gauge invariants from these:
\Eqrsub{
&& \h F= F -\frac{S_0 r}{k} D^r\sigma_T - \frac{kS_0}{2} \sigma_T,\\
&& \h F_{ab}=F_{ab} -\frac{S_0}{k}\inrbra{D_a(r^2D_b(\sigma_T))+D_b(r^2D_a(\sigma_T))},\\
&& \tilde \sigma^t=\h \sigma^t - \frac{S_0 r^2}{\mu k} D^t(\sigma_T),\\
&& \tilde \sigma^r=\h \sigma^r - \frac{r^2}{k} D^r(\sigma_T).
}
The perturbation equations for these variables are obtained by the replacements
\Eq{
F\tend \h F,\quad
F_{ab}\tend \h F_{ab},\quad
\h \sigma^a \tend \tilde\sigma^a,\quad
\h\sigma_T\tend 0.
}
Hence, the general solution for this case is expressed in terms of these variables as
\Eqrsub{
&& \h F=-\frac{k}{2}L-\frac{r}{k}D^r L,\\
&& \h F_{ab}=-\frac1{k^2}\inrbra{D_a(r^2D_b L)+D_b (r^2D_aL)},\\
&& \tilde \sigma^a=0,
}
where $L(t,r)$ is an arbitrary function.

Finally, for the exceptional modes with $l=0$, from the gauge transformation formula
\Eqrsub{
&& \delta_g f_{ab}=-D_aT_b - D_b T_a,\\
&& \delta_g H_L= -\frac1{r}T^r,
}
we can construct the following gauge invariants from $f_{ab}$, $H_L$ and $\sigma_a$:
\Eqrsub{
&& \h F_{ab}= f_{ab}- D_a \tilde\sigma_b - D_b \tilde\sigma_a,\\
&& \h F = H_L - \frac{S_0}{r}\sigma^r,
}
where
\Eq{
\tilde\sigma^t= \sigma^t,\quad
\tilde\sigma^r=S_0 \sigma^r.
}
We have shown that the general solution for $l=0$ can be expressed in terms of these gauge invariants as
\Eqrsub{
&& \h F_{tt}=\delta M \pd_M g_{tt} + 2f \d T^t,\\
&& \h F_{tr}=\delta M \pd_M g_{tr} +  f(h' \d T^t + \pd_r T^t)-f' T^t,\\
&& \h F_{rr}= \delta M\pd_M g_{rr} + 2h' f\pd_r T^t,\\
&& \h F=0,
}
where $T^t$ is an arbitrary function of $t$ and $r$, and $\delta M$ is an arbitrary constant corresponding to the mass variation.

\section{Summary and Conclusions}   \label{sec:Conc}

In the present paper, we first looked for the parameter relation for which the non-linear massive gravity theory admits the Schwarzschild-de Sitter black hole as an exact solution systematically. We found that when the parameters satisfies the relation $\beta=\alpha^2$, there exists a family of solutions parameterized by an arbitrary function $T_0(t,r)$, which are isomorphic to the Schwarzschild-de Sitter spacetime but are not equivalent if the configuration of the St\"uckelberg fields are taken into account.

We next investigated the perturbative stability of this family of  Schwarzschild-de Sitter-type black holes in the framework of the dRGT formulation of the non-linear massive gravity with $\beta=\alpha^2$. We found that the perturbative equations derived from the field equations of the dRGT theory becomes identical to the perturbations equation for the vacuum Einstein theory with cosmological constant if we take into account the consistency condition obtained from the field equations by the Bianchi identity. This consistency condition is essentially equivalent to the field equation for the St\"uckelberg field. This implies that the Schwarzschild-de Sitter black hole solution is stable in the non-linear massive gravity theory
as far as the spacetime structure is concerned
at least in the linear perturbation level, in contrast to the bi-Schwarzschild solution in the bi-metric theory.

In spite of this stability result, we found a pathological feature of the black hole solution in the dRGT theory with the parameter relation $\beta=\alpha^2$; the general solution to the perturbation equations contains an arbitrary function of the spacetime coordinates. This implies that the predictability of dynamics is lost at least in the linear perturbation level around this black hole solution. This degeneracy can be removed by coordinate transformations if we neglect the St\"uckelberg fields. Hence, the pathology appears to come from the dynamics of the St\"uckelberg fields. Because the Schwarzschild-de Sitter black hole becomes an exact solution only when the higher-order mass terms exist, there is a possibility that this pathology might be removed in the non-linear level of perturbations.

\section*{Acknowledgement}

This work is supported by  the Grant-in-Aid for Scientific Research
(A) (22244030) and the Grant-in-Aid for Scientific Research on
Innovative Area No. 21111006. I.A would like to thank Gregory Gabadadze, Ruth Gregory and Robert Brandenberger for useful comments and suggestions during PASCOS 2013 organized at National Taiwan University, Taipei. I.A also would like to thank Takahiro Tanaka for useful discussions and comments during the JGRG23 organized at Hirosaki University.

\appendix

\section{The other parameter choices admitting the Schwarzschild-de Sitter solution}
\label{App:TheOtherCases}

In this appendix, we exhaust all possible choices of the parameters in which $X$ becomes a constant multiple of the unit matrix as $m^2 X^\mu_\nu=\Lambda\delta^\mu_\nu$ assuming that the spacetime metric takes the spherically symmetric form \eqref{SSmetric} and the St\"uckelberg fields satisfy the unitary gauge condition \eqref{UnitaryGauge}. We use the same notations as in \S III.

\Bitm
\item[i)] $c=0, a=b$.  In this case, the condition
\Eq{
X^t_t-X^\theta_\theta=(F_1+a F_2)\inpare{a-1+\frac{1}S}=0,
}
implies $a=1-1/S$ or $a=-F_1/F_2$, if we exclude the case $F_1=F_2=0$ discussed in Section \ref{sec:BHsol}.  Then, from $X^t_t=\Lambda/m^2=\const$, if follows that $S$ is constant. Hence, the metric must represent a flat spacetime and $\Lambda=0$. Because the metric \eqref{SSmetric} with constant coefficients has vanishing curvature if
\Eq{
g_{(2)}=S^2 g_{tt} \equivalent
S^2\inrbra{(1-b)^2-c^2}=1
\label{flatmetric:regularity}
}
in general, in the present case, we obtain the constraint $a=b=1-1/S$. The corresponding flat metric should have the form
\Eq{
ds^2=S^2(-dt^2+dr^2+r^2d\Omega^2).
\label{Sol:F:metric}
}
No constraint on $\alpha$ and $\beta$ is required, but the value of $S$ is restricted from the condition $X^t_t=0$ to
\Eq{
S=1,\ \frac{3\alpha+2\beta\pm\sqrt{9\alpha^2-12\beta}}{2(3+3\alpha+\beta)}.
\label{Sol:F:S}
}
The case $a=-F_1/F_2$ can be included in this solution as a special case.
\item[ii)] $c=0,a\neq b$.  In this case, we obtain  $F_3=0$ from $X^t_t=X^r_r$, hence $S$ must be constant. From this it follows that $X^t_t=(1/S-1)(F_1+1)=\const$. Next, from
\Eq{
0=X^t_t-X^\theta_\theta= (F_1+bF_2)(a-1+1/S),
}
we obtain $a=1-1/S$ or $b=-F_1/F_2$, if we exclude the case $F_1=F_2=0$ discussed Section \ref{sec:BHsol}.  Because $c=0$, $T_0$ should be a function only of $t$ from \eqref{SdSsol:general}. Hence, if $a=1-1/S=const$, the metric should be flat because $g_{tt}$ is constant from \eqref{gbyabc}. Then, from \eqref{flatmetric:regularity}, we obtain b=1-1/S=a, contradicting the assumption. Next, when $b=-F_1/F_2=\const$, we find that the metric is flat and $a$ is constant again.  Now, from $X^t_t=-(F_1+1)(1-1/S)=0$we obtain two constraints $F_1=-1, F_2=S/(S-1)$ because $S=1$ leads to $F_3=1$. This means that $b=1-1/S$. This leads to the contradiction due to the regularity condition \eqref{flatmetric:regularity}. Thus, this case has no other solution than those discussed in Section \ref{sec:BHsol}.
\item[iii)] $c\neq0, a=b$. In this case, $F_3=0$ is required, and from $X^t_t=\const$, it follows that $S$ is constant. Then, from
\Eq{
0=X^t_t-X^\theta_\theta=F_2 \inrbra{\inpare{a-1+1/S}^2+c^2},
}
we obtain $F_2=0$. Hence, this case is a special case of the case with  $F_1=F_2=0$ discussed in Section \ref{sec:BHsol}.

\item[iv)] $c\neq0,a\neq b$. Again, we obtain $F_3=0$ and $S=\const$. If $F_2=0$, this case reduces to the class $F_1=F_2=0$ discussed in Section \ref{sec:BHsol}. Next, when $F_2\neq0$, the constraint
\Eq{
0=X^t_t-X^\theta_\theta = F_2\insbra{{(a-1+1/S)(b-1+1/S)+c^2}}
}
leads to
\Eq{
0=c^2+(1-a-1/S)(1-b-1/S)=c^2+(1-a)(1-b)-\frac{2-a-b}{S}+\frac{1}{S^2}.
\label{Constraint:case3}
}
This should be satisfied by $a$, $b$, $c$ corresponding to the Schwarzschild-de Sitter metric
\Eq{
ds^2=-f(Sr)dT_0(t,r)^2+\frac{S^2dr^2}{f(Sr)}+S^2r^2d\Omega^2.
\label{Sol:SdS:metric}
}
From the general formula in Section \ref{sec:BHsol}, we obtain
\Eqrsub{
&& c^2+(1-a)(1-b)=\frac{1}{S|\d T_0|},\\
&& 2-a-b=M_1=\frac{1}
 {S|\dot{T}_0|}\left(f\dot{T}_0^2+\frac{S^2}{f}-f(T'_0)^2+2S\dot{T}_0\right)^{1/2},
}
where $\d T_0=\pd_t T_0$, $T_0'=\pd_r T_0$, and $f=f(Sr)$ is understood. Inserting these into the above constraint, we obtain
\Eq{
(T_0')^2 =\frac{1-f(Sr)}{f(Sr)}\inpare{\frac{S^2}{f(Sr)}-\d T_0^2}.
\label{Sol:SdS-II:T0:Constraint}
}
Hence, in this case, no relation is imposed on $\alpha$ and $\beta$, but instead the gauge transformation function $T_0(t,r)$ is constrained. The value of $S$ is determined by $F_3=0$ as
\Eq{
S=\frac{\alpha+\beta\pm\sqrt{\alpha^2-\beta}}{1+2\alpha+\beta},
\label{Sol:SdS-II:S}
}
and the corresponding cosmological constant is given by
\Eq{
\Lambda = -m^2 \inpare{1-\frac{1}{S}}\inpare{2+\alpha -\frac{\alpha}{S}}.
\label{Sol:SdS-II:Lambda}
}
The condition $\Lambda\neq0$ is given by
\Eq{
\beta\neq \frac34 \alpha^2 \equivalent \Lambda\neq0.
}

Note that \eqref{Sol:SdS-II:T0:Constraint} has a solution for $T_0$ locally with respect to $r$ at most in general. One exception is the solution
\Eq{
T_0= St \pm \int^{Sr}\inpare{ \frac{1}{f(u)}-1}du.
}
Interestingly, this corresponds to a Finkelstein-type time coordinate which is regular at the future horizon or the past horizon.
\Eitm

Finally, as the summary of this appendix, we give an exhaustive list of the spherically symmetric solutions isomorphic to the Schwarzschild-de Sitter solution and the corresponding parameter constraints in the dRGT massive gravity theory:

\Bitm
\item {\bf Solution F}:  The solution with a flat metric.  The metric form should be that of \eqref{Sol:F:metric} with $S$ given by one of the values in \eqref{Sol:F:S}. No constraint on the parameters $\alpha$ and $\beta$ is required.
\item {\bf Solution SdS-I}: The Schwarzschild-de Sitter type solution discussed in Section \ref{sec:BHsol}. The cosmological constant is given by $\Lambda=m^2/\alpha$, and the metric is given by \eqref{Sol:SdS:metric} with $S=\alpha/(1+\alpha)$. The parameters are constrained as $\beta=\alpha^2$, but the function $T_0(t,r)$ can be arbitrary.
\item {\bf Solution SdS-II}: The Schwarzschild-de Sitter type solution whose metric is given by \eqref{Sol:SdS:metric} with constant $S$ given by \eqref{Sol:SdS-II:S} and the cosmological constant \eqref{Sol:SdS-II:Lambda}.  The parameters $\alpha$ and $\beta$ are weakly constrained as $\beta < \alpha^2$, but the function $T_0$ is constrained to those satisfying \eqref{Sol:SdS-II:T0:Constraint}.
\Eitm

\section{Explicit forms for $\delta Q_1$, $\delta Q_2$ and $\delta Q_3$} \label{App:delQn}

\Eqr{
\delta Q_1 &=& \frac{2(\alpha+1)}{\alpha}H_L Y+ \frac{1}{2}\inrbra{(a-1)^3-c^2(2a+b-3)} h_{tt}
\notag\\
 && + \frac{\alpha^2 }{2(\alpha+1)^2}\inrbra{-(b-1)^3+ c^2(a+2b-3)}h_{rr}
 \notag\\
 && + \frac{\alpha c}{\alpha+1}\inrbra{c^2-(a^2+b^2+ab-3a-3b+3)}h_{tr},
}
\Eqr{
\delta Q_2 &=& -\frac{4(\alpha+1)}{\alpha^2}H_L Y
 + \inrbra{c^4-(3a^2+b^2+2ab-6a-3b+3)c^2+a(a-1)^3}h_{tt}
 \notag\\
 && + \frac{\alpha^2}{(\alpha+1)^2}\inrbra{-c^4+c^2(a^2+3b^2+2ab-3a-6b+3)-b(b-1)^3}h_{rr}
 \notag\\
 && +\frac{2\alpha c}{\alpha+1}\inrbra{c^2(2a+2b-3)-a^3-b^3-ab(a+b)+3a^2+3b^2+3ab-3a-3b+1}h_{tr},
 \notag\\
 &&
}
\Eqr{
\delta Q_3&=& \frac{6(\alpha+1)}{\alpha^3}H_L Y
+\frac{3}{2}\Big\{ c^4(3a+2b-3)
 \notag\\
&& -c^2(4a^3+b^3+2ab^2+3a^2b-9a^2-3b^2-6ab+6a+3b-1)+a^2(a-1)^3\Big\}h_{tt}
\notag\\
&& -\frac{3\alpha^2}{2(\alpha+1)^2}\Big\{ c^4(3b+2a-3)
 \notag\\
&&\qquad -c^2(4b^3+a^3+2ba^2+3b^2a-6ab -9b^2-3a^2+6b+3a-1)+b^2(b-1)^3\Big\}h_{rr}
\notag\\
&& + \frac{3\alpha c}{\alpha+1}\Big\{-c^4+c^2(3a^2+3b^2+4ab-6a-6b+3)
 -a^4 - b^4 -ab^3-a^3b -a^2b^2
 \notag\\
&&\quad
 +3a^3 +3b^3+3ab^2+3a^2b -3a^2-3b^2-3ab+a+b \Big\} h_{tr}.
}

\vfill\pagebreak

\end{document}